\documentclass[
 reprint,
 amsmath,amssymb,
 pra,
 superscriptaddress,
]{revtex4-1}
\usepackage{graphicx}
\usepackage{dcolumn}
\usepackage{bm}
\usepackage{color}

\begin{document}

\title{Bloch oscillations in arrays of helical waveguides}

\author{WeiFeng Zhang}
\affiliation{School of Physics and Astronomy, Shanghai Jiao Tong University, Shanghai 200240, China}
\author{Xiao Zhang}
\affiliation{School of Physics and Astronomy, Shanghai Jiao Tong University, Shanghai 200240, China}
\affiliation{Department of Physics, Zhejiang Normal University, Jinhua 321004, China}
\author{Yaroslav V. Kartashov}
\affiliation{ICFO-Institut de Ciencies Fotoniques, The Barcelona Institute of Science and Technology, 08860 Castelldefels (Barcelona), Spain}
\affiliation{Institute of Spectroscopy, Russian Academy of Sciences, Troitsk, Moscow 108840, Russian Federation}
\author{Xianfeng Chen}
\affiliation{School of Physics and Astronomy, Shanghai Jiao Tong University, Shanghai 200240, China}
\author{Fangwei Ye}
\email{fangweiye@sjtu.edu.cn}
\affiliation{School of Physics and Astronomy, Shanghai Jiao Tong University, Shanghai 200240, China}

\begin{abstract}
We study optical Bloch oscillations in the one- and two-dimensional arrays of helical waveguides with transverse refractive index gradient. Longitudinal rotation of waveguides may lead to notable variations of the width of the band of quasi-energies and even its complete collapse for certain radii of the helix. This drastically affects the amplitude and direction of Bloch oscillations. Thus, they can be completely arrested for certain helix radii or their direction can be reversed. If the array of helical waveguides is truncated and near-surface waveguide is excited, helix radius determines whether periodic Bloch oscillations persist or replaced by the irregular near-surface oscillations.
\end{abstract}

\maketitle

\section{Introduction}

Bloch oscillations (BOs) is a famous physical phenomenon manifested as time-periodic evolution of a wavepacket in a spatially-periodic potential in the presence of transverse potential gradient (force). Physically the emergence of BOs is connected with appearance of equidistant spectrum with localized eigenmodes in the presence of potential gradient. Introduced for the first time for electrons moving in a crystal under the action of a constant electric field \cite{Bloch1,Bloch2}, BOs were observed for electrons in semiconductor superlattices \cite{semiconductor1,semiconductor2}, shortly after observation in them of the Wannier-Stark ladder \cite{ladder1,ladder2}. As a universal wave phenomenon BOs were demonstrated in a variety of physical systems, including ultracold atoms \cite{ultracold atoms1,ultracold atoms2}, Bose-Einstein condensates held in optical lattices \cite{condensates1,condensates2}, waveguide arrays \cite{ol1998,prl1999_1,prl1999_2} or optically-induced lattices \cite{kivshar2006}, surface plasmon waves in plasmonic crystals \cite{nc2014}, and parity-time symmetric systems \cite{parity-time}. Waveguide arrays allow observation of unusual types of BOs, including fractional oscillations \cite{complex1,complex2}.

Optical BOs are most frequently considered in periodic structures with constant transverse refractive index gradient, such as effective gradient induced by circular waveguide bending \cite{lenz1999}. Nevertheless, even small periodic longitudinal modulations of the parameters of waveguides may strongly affect coupling between them that, in turn, changes the entire dynamics of light propagation. The progress in research in this direction is summarized in recent review \cite{review}. Examples of rich possibilities for control of light propagation arising due to periodic longitudinal modulations of guiding structures include diffraction management in zigzag arrays \cite{diffm}, dynamic localization in periodically curved arrays \cite{dl}, inhibition of tunneling \cite{inhibit}, Rabi oscillations \cite{rabi1,rabi2}, and topological effects in arrays with helical waveguides \cite{moti}. Under appropriate conditions periodic longitudinal modulations may induce dynamic band collapse, as suggested in honeycomb \cite{njp} and rhombic \cite{mukh1} lattices. Since such collapse leads to qualitative modification of the spectrum of the system it may drastically affect BOs. Nevertheless, the effect of dynamic band collapse caused by longitudinal modulation on BOs remains largely unexplored, since many works utilized only one type of modulation that leads either to constant or time-periodic potential gradient in the system possessing flat bands even in the absence of modulation \cite{mukh1,flach}.

In this paper we study the interplay between BOs and dynamic band collapse in a simple one- or two-dimensional array of helical waveguides. When helix radius is zero, the bands of such arrays are dispersive. We show that dynamic band collapse taking place for certain helix radii is accompanied by suppression of BOs. Each time when helix radius passes the value at which band collapse occurs, the direction of BOs is inverted. We also study BOs in truncated one-dimensional waveguide arrays.
\begin{figure}[htp]
\centering
{\includegraphics[width=1\linewidth]{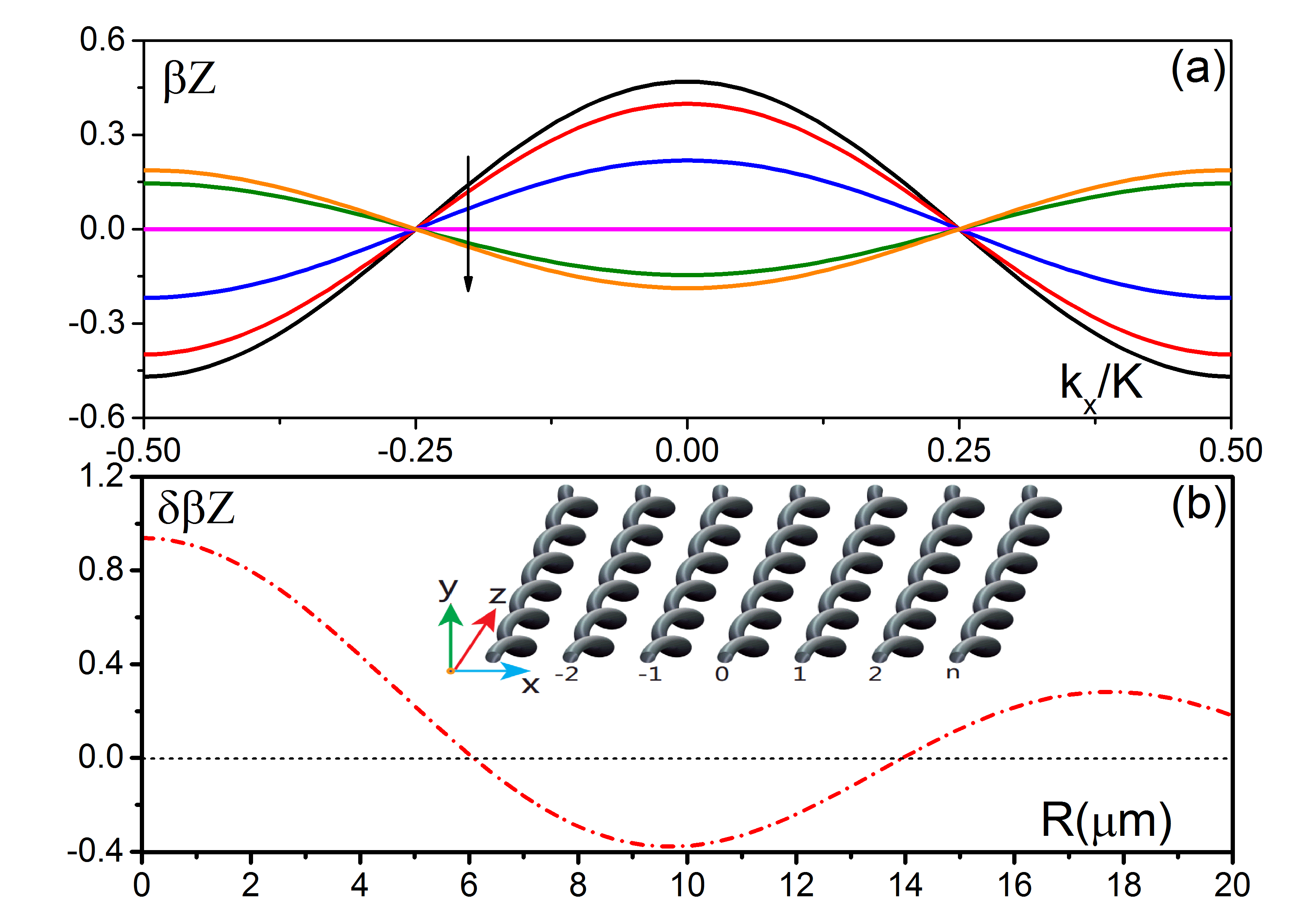}}
\caption{ (a) $\beta Z$ versus normalized Bloch momentum $k_x/\textrm{K}$ (dispersion curves) for the 1D array array of helical waveguides for different radii of the helix $R=0 ~\mu\textrm{m}$, $2 ~\mu \text{m}, 4~\mu \text{m}, 6.05~\mu \text{m}, 8~\mu \text{m}$, and $10 \mu \text{m}$. Arrow indicates the direction of increase of helix radius $R$. (b) Scaled difference $\delta \beta Z$ of quasi-energies in the center and at the edge of the Brillouin zone versus helix radius $R$. The inset shows schematic illustration of the 1D waveguide array.}
\label{fig:1}
\end{figure}
\section{Bloch oscillation in one-dimensional helical waveguides}

\label{sec:2}

We start our analysis by considering paraxial propagation of light along the $z$ axis in modulated waveguide array that can be described by the following Schr\"{o}dinger equation:
\begin{equation}
\begin{gathered}
    i\frac{{\partial \psi}}{{\partial z}}{\text{ = }}-\frac{1}{{{\text{2}}{k_0}}}\left( {\frac{{{\partial ^{\text{2}}}}}{{\partial {x^2}}}{\text{ + }}\frac{{{\partial ^{\text{2}}}}}{{\partial {y^2}}}} \right)\psi {\text{}} - \frac{{{k_0}}}{{{n_0}}}[\Delta n(x,y,z)+\alpha x]\psi {\text{}}.
\end{gathered}
\label{eq:refname1}
\end{equation}
Here $\psi (x,y,z)$ is the envelope of the electric field $E(x,y,z) = \psi(x,y,z)\exp (i{k_0}z-i\omega t)$; $k_0 =2\pi n_0/\lambda$ is the wavenumber in the material with refractive index ${n_0}$; $\omega  =2\pi c/\lambda$ is the frequency; $\lambda $ is the wavelength. The function $\Delta n(x,y,z)$ describes refractive index distribution in a waveguide array consisting of helical waveguides with helix period $Z$ and radius $R$ [see the inset in Fig.~1(b) showing example of one-dimensional array]; the separation between waveguides is $d$. The parameter $\alpha>0$ stands for the transverse refractive index gradient in the $x$ direction, that is required for the occurrence of BOs. Eq.~(1) can be rewritten in the coordinate frame $x'=x+R\cos (\Omega z)$ and $y'=y+R\sin (\Omega z)$ co-rotating with the waveguides, where $\Omega  = 2\pi /Z$ is the rotation frequency:
\begin{equation}
 i\frac{{\partial \psi }}{{\partial z}}{\text{=}}-\frac{1}{{{\text{2}}{k_0}}}[{{\nabla+i\textbf{A}(z)}]^2}\psi-\frac{{{k_0}{R^2}{\Omega ^2}}}{2}\psi-\frac{{{k_0}}}{{{n_0}}}[\Delta n(x,y)+\alpha x]\psi.
 \end{equation}
Here $\nabla=(\partial_{x},\partial_{y})$;  $\textbf{A}(z) = {k_0}R\Omega [\sin (\Omega z), - \cos (\Omega z)]$ is an effective gauge potential arising due to waveguide rotation and proportional to the radius of helix $R$ (we have omitted primes in coordinates here). In this new coordinate frame the waveguides are straight and refractive index profile is described by a $z$-independent function $\Delta n(x,y)=p\sum_m \exp[-[(x-md)^2+y^2]^8/a^{16}]$, where $p$ is the refractive index modulation depth in each waveguide, and $a$ is the waveguide width. We further use parameters $p=7\times 10^{-4}$,  $a=3.8~\mu \text{m}$, $n_0=1.45$, and $d=17.5~\mu\text{m}$ typical for helical waveguide arrays that can created using developed technology of fs-laser writing \cite{moti}. For a wavelength $\lambda=633~\text{nm}$, isolated waveguide with these parameters supports only one guided mode with propagation constant $\beta_0/k_0 \approx1.34 \times 10^{-4}$.

As mentioned above, helical waveguide arrays can be created using the developed fs-laser writing technology~\cite{moti}. The index gradient term of Eq.~(2), $\alpha x$, that is required for the occurrence of Bloch oscillation can be introduced by bending of helical array as a whole along parabolic trajectory~\cite{alex_bending}. With this scheme, the index gradient $\alpha$ is inversely proportional to the curvature of the bending. Another way to introduce the index ramp across the waveguide array is to use waveguide arrays based on thermo-optic polymers. Notice that tunable optical Bloch oscillations have already been observed in thermo-optic polymer arrays by applying a temperature gradient across the waveguide array~\cite{prl1999_2}. Thus, fs-laser writing technology can potentially be used to write helical waveguides in such polymers, while control of the refractive index gradient can be achieved by varying temperature at the opposite sides of the array (the latter implies possibility of slow, but dynamical variation of the gradient).

Further we employ the standard tight-binding approximation, assuming coupling only between nearest waveguides in the array. We also assume that helix period $Z=0.4~\textrm{cm}$ used here exceeds Rayleigh length, so that radiative losses are low. Using tight-binding approximation and Pierls substitution \cite{pierls}, one obtains for one-dimensional array the following coupled-mode equations for the field amplitude ${\psi _n}(z)$ in the $n$-th waveguide:
\begin{equation}
 i\frac{{\partial {\psi _n}(z)}}{{\partial z}} =\sum_{m=n\pm1} {c{e^{i[\textbf{A}(z) \cdot {\textbf{r}_{mn}} + ({\beta _m} - {\beta _n})z]}}} {\psi _m}(z),
\label{eq:refname1}
\end{equation}
where $c=60.3~\text{m}^{-1}$ is the coupling constant between neighboring (straight) waveguides evaluated for parameters of our array, ${\textbf{r}_{mn}}$ is the displacement vector between waveguides $m$ and $n$, and $\beta _n=\beta_0+\alpha k_0 dn/n_0$. For the gradient $\alpha=0.2 ~\text{m}^\text{-1}$ one obtains $\alpha k_0 d/n_0 \approx 34.7~\textrm{m}^{-1}$.

\begin{figure}[htp]
\centering
{\includegraphics[width=1\linewidth]{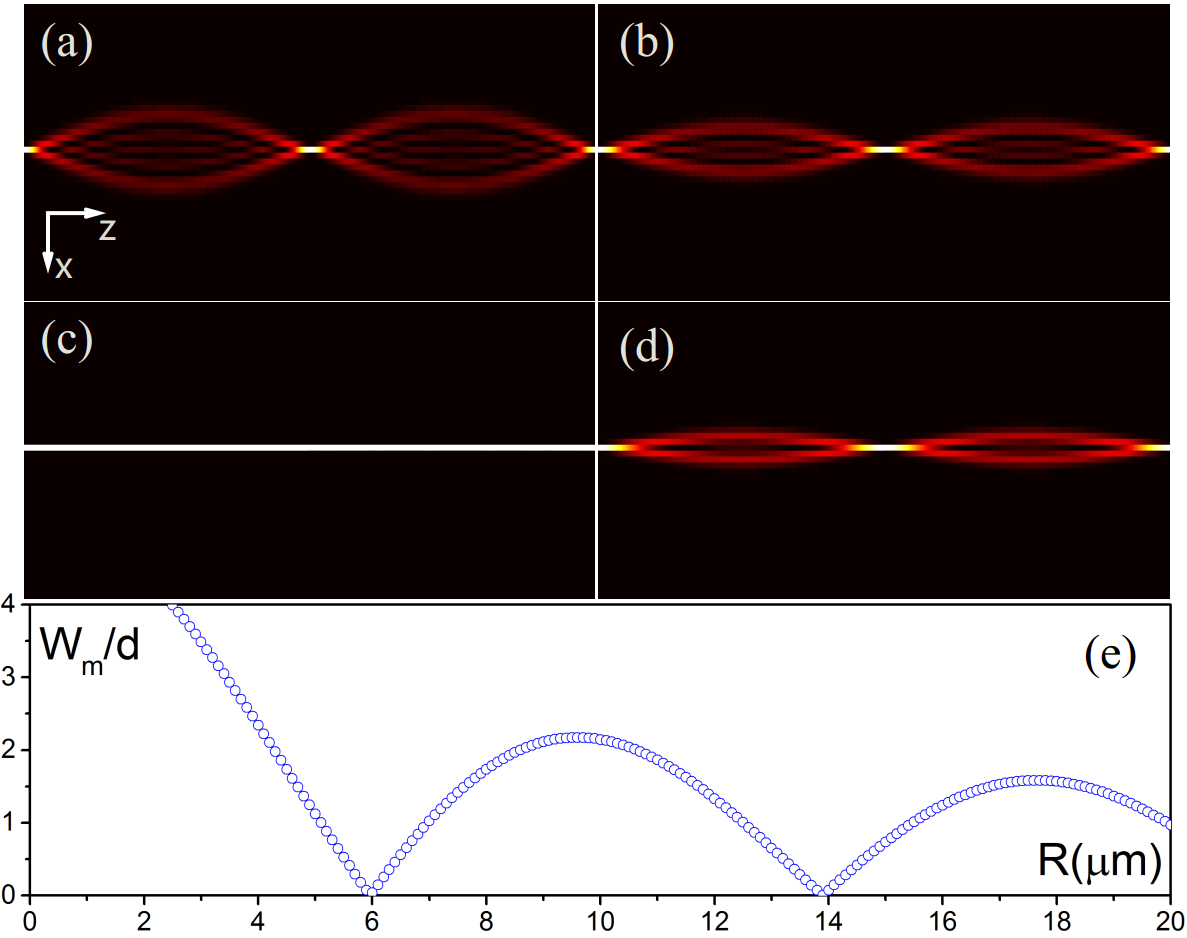}}
\caption{ Dynamics of light propagation in the 1D array of helical waveguides with a transverse refractive index gradient $\alpha=0.2~\textrm{m}^{-1}$, when only one central waveguide is excited (narrow excitation). Helix radius radius $R=0 ~\mu \textrm{m}$ (a), $3~ \mu \text{m}$ (b), $6.05 ~\mu \text{m}$ (c), and $10~ \mu \text{m}$ (d). (e) Maximal scaled width $\text{W}_\textrm{m}/d$ of the wavepacket, acquired upon propagation, versus helix radius $R$.}
\label{fig:2}
\end{figure}

First we address the impact of waveguide rotation on the band structure of one-dimensional waveguide array and consider its eigenmodes at $\alpha=0$ (i.e. $\beta_m-\beta_n=0$) in Eq.~(3). Since right-hand side of Eq. (3) is $z$-dependent, static eigenmodes do not exist, instead, solutions have the form of Floquet modes ${\psi_n}(z)=\exp (i\beta z+ik_x nd){\varphi _n}(z)$, where ${\varphi _n}(z)$ are $z$-periodic functions with a period equal to helix period $Z$, and $\beta$ is the quasi-energy. The dependencies of $\beta$ on normalized momentum $k_x/\textrm{K}$, where $\textrm{K}=2\pi/d$ is the width of the Brillouin zone, are depicted in Fig.~1(a) for different helix radii $R$. One can see that the width of the band strongly depends on helix radius $R$. While in the absence of rotation one observes usual dependence with $\beta|_{k_x=0}>\beta|_{k_x=\textrm{K}/2}$, the band experience dynamic collapse around $R\approx 6.05~\mu \text{m}$ and becomes completely flat in the tight-binding approximation. With further increase of helix radius $R$ the curvature of the band is inverted and one obtains the dependence with $\beta|_{k_x=0}<\beta|_{k_x=\textrm{K}/2}$, but afterwards another band collapse occurs, i.e. the effect repeats as $R$ increases. This is clear from the dependence of difference of quasi-energies $\delta \beta=\beta|_{k_x=0}-\beta|_{k_x=\textrm{K}/2}$ in the center and at the edge of the Brillouin zone, on helix radius $R$ shown in Fig.~1(b), where critical values of helix radius $R_\textrm{cr}$ corresponding to zero $\delta\beta$ can be identified. Band collapse is naturally connected with renormalization of coupling constant $c_\textrm{eff}=c \mathcal{J}_0(k_0 R \Omega d)$ caused by waveguide rotation by analogy with renormalization caused by sinusoidal driving \cite{review}. Critical values $R_\textrm{cr}$ correspond to zeros of $c_\textrm{eff}$.

To study the impact of dynamic band collapse on BOs we now assume nonzero gradient $\alpha \neq 0$ in Eq.~(3), i.e. nonzero difference of propagation constants $\beta_m-\beta_n$. First, we address the case of narrow excitation of the central $(n=0)$ waveguide of the 1D array, i.e. $\psi_{n=0}=1$ and $\psi_{n\neq 0}=0$ at $z=0$. Dynamics of light propagation $|\psi_n (z)|$ for single-waveguide excitation is shown in Fig.~2 for various helix radii $R$. For straight waveguides [$R=0~\mu \textrm{m}$, Fig.~2(a)], the input excitation strongly expands, but then shrinks and completely restores the input distribution after each period $Z_\text{Bloch}=\lambda/(\alpha d)$ of BOs. As it was mentioned above, this restoration is a consequence of formation of equidistant spectrum of localized modes with the difference between neighboring eigenvalues equal to $2\pi \alpha d/\lambda$. When waveguides are made helical one still observes periodic expansion and shrinkage of the beam with the same $z$-period, but the width of the wavepacket in the point of its maximal expansion $z=Z_\textrm{Bloch}/2$ gradually reduces with increase of $R$ [Fig.~2(b)]. BOs are completely suppressed when radius of helix approaches critical value $R=R_{\text{cr}}$ corresponding to band collapse [Fig.~2(c)]. In this case, despite the presence of small transverse gradient the excitations with all momenta $k_x$ acquire the same phase shifts upon propagation and therefore do not diffract. Further increase of helix radius $R$ results in restoration of BOs, but with substantially smaller maximal expansion of the wavepacket [Fig.~2(d)]. The influence of the helix radius $R$ on the maximal width of the wavepacket $W_\textrm{m}/d$ in the course of BOs, acquired at the distance $z=Z_\text{Bloch}/2$, is illustrated in Fig.~2(e). Notice that this curve qualitatively reproduces the dependence of the width of quasi-energy band $|\delta\beta Z|$ on helix radius from [Fig.~1(b)].

The rotation of waveguides in the array affects not only amplitude of the BOs, but it can also invert their direction. To illustrate this one has to consider evolution of broad excitations that are known to exhibit transverse shifts during BOs, rather than width oscillations. Figure 3 illustrates dynamics of BOs for inputs in the form of sufficiently broad Gaussians $\psi_{n,z=0}=\text{exp}[-(n/w)^2]$, where $w=4$. In the unmodulated system such waveguide oscillates periodically in the transverse plane, returning at $z=Z_\textrm{Bloch}$ to its input location [Fig. 3(a)]. Oscillations occur in the region $n>0$ in accordance with positive refractive index gradient $\alpha>0$.  As in the case of narrow excitations, the amplitude of oscillations of the wavepacket center decreases with increase of helix radius $R$ [Fig. 3(b)], and at $R=R_{\text{cr}}$ the oscillations are completely arrested [Fig. 3(c)]. Subsequent reappearance of BO at $R>R_\textrm{cr}$ is accompanied by the reversal of the direction of oscillations [Fig. 3(d)] - a counterintuitive effect, taking into account the facts that such oscillations occur in the direction opposite to the refractive index gradient and that initial excitation had trivial phase distribution. Overall scaled displacement $D_\textrm{m}/d$ of the wavepacket center calculated at the distance $z=Z_\text{bloch}/2$ is presented in Fig.~3(e) as a function of helix radius $R$. This dependence closely matches the dependence of the difference of quasi-energies $\delta \beta=\beta|_{k_x=0}-\beta|_{k_x=\textrm{K}/2}$ on $R$ from Fig. 1(b). Thus, even though there are no static eigenmodes in helical array, one can still assume that by analogy with BOs in static arrays in our case the initial excitation with narrow spectrum in the $k$-space moves across the Brillouin zone [Fig. 1(a)] under the action of small $(\alpha d <<1)$ refractive index gradient, and that position of its center in the real space is approximately described by the formula, analogous to that describing BOs in continuous systems \cite{Bloch1}:
\begin{equation}
D(z)=D_0+(n_0/\alpha k_0)[\beta(k_{x0})-\beta(k_{x0}-\textrm{K}\alpha dz/\lambda)]
\label{eq:refname1}
\end{equation}
where $k_{x0}$ is the initial momentum and $D_0$ is the initial displacement at $z=0$. This formula indeed gives very good agreement with numerically calculated dependence shown in Fig. 3(e). In particular, the points where displacement is zero, clearly corresponds to $R$ values at which $\beta(k_x)=\textrm{const}$.

\begin{figure}[htp]
\centering
{\includegraphics[width=1\linewidth]{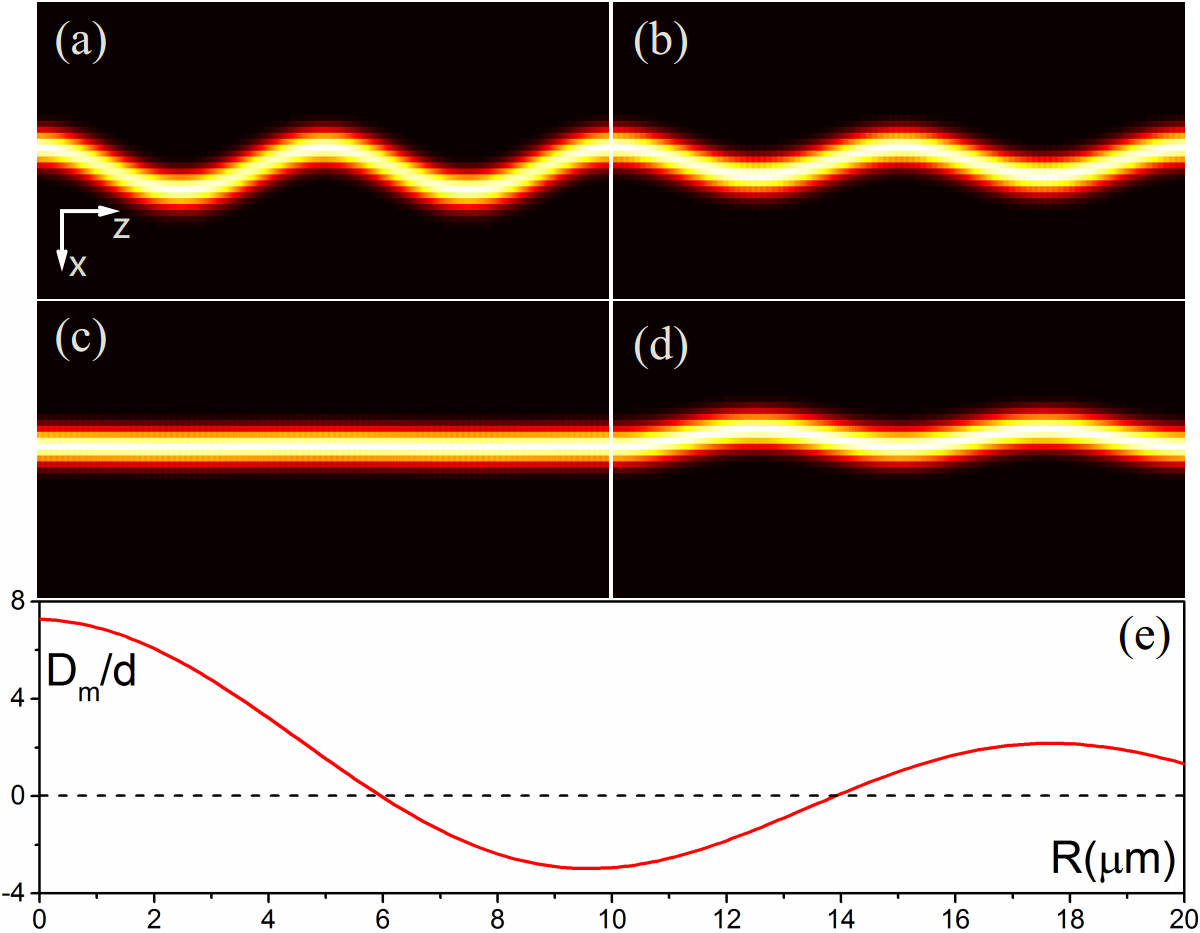}}
\caption{ Dynamics of light propagation in the 1D array of helical waveguides with a transverse refractive index gradient $\alpha=0.2$, when multiple waveguides are excited by a broad Gaussian beam. Helix radius $R=0 ~\mu \textrm{m}$ (a), $3~ \mu \textrm{m}$ (b), $6.05 ~\mu \textrm{m}$ (c), and $10~ \mu \textrm{m}$ (d), respectively. (e) Maximal scaled transverse displacement of the wavepacket center $\text{D}_\textrm{m}/d$, acquired upon propagation, versus helix radius $R$.}
\label{fig:3}
\end{figure}

The inversion of the direction of BOs for $R<R_\textrm{cr}$ and $R>R_\textrm{cr}$ implies qualitatively different dynamics when a Gaussian beam is launched near the boundary of truncated helical array. The propagation dynamics for such near-surface excitation is presented in Fig.~4, where we used the input wavepacket $\psi_{n,z=0}=\text{exp}[-(n-n_c)^2/w^2]$ of width $w=4$ for $n_c=25$, where array is truncated on the waveguide with $n=25$ that has the highest refractive index. BOs do not occur for $R<R_\textrm{cr}$ [Figs.~4(a) and 4(b)]. Instead, the wavepacket experiences generally aperiodic near-surface oscillations because gradient stimulates displacement toward the surface that is compensated by repulsion from the surface. At  $R=R_\textrm{cr}$ one observes formation of stationary linear surface wave [Fig. 4(c)]. Remnants of BOs are observable only for $R>R_\textrm{cr}$, when the wavepacket experiences displacement against gradient and moves away from the surface of array. In this case the wavepacket periodically returns to the surface, the oscillations are nearly periodic [Fig. 4(d)].

\begin{figure}[htp]
\centering
{\includegraphics[width=1\linewidth]{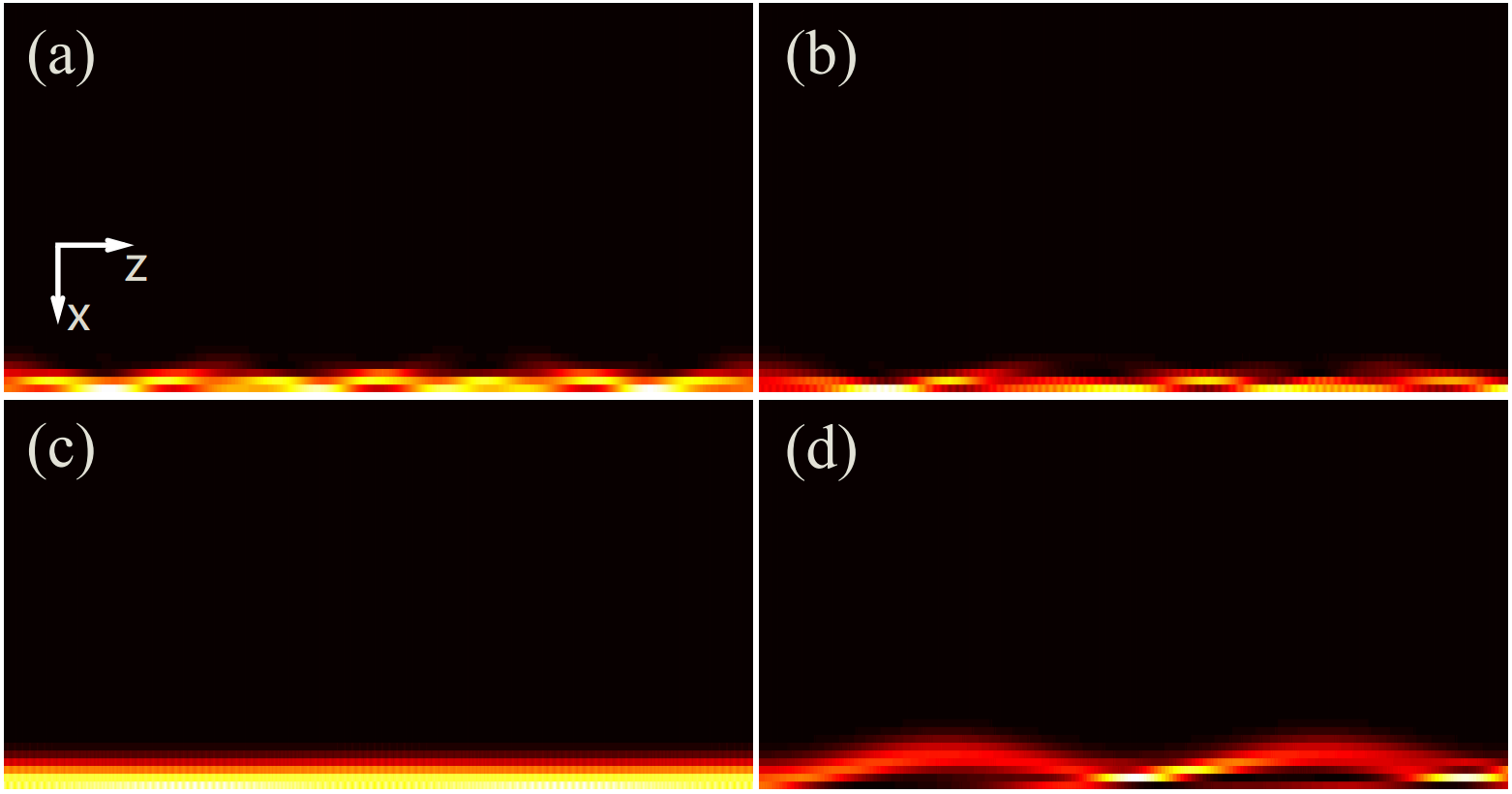}}
\caption{ Propagation dynamics around the edge of truncated 1D array of helical waveguides with a transverse refractive index gradient $\alpha=0.2$ for $R=0~\mu \textrm{m}$ (a), $3~ \mu \text{m}$(b), $6.05~\mu \text{m}$ (c), and $10~\mu \text{m}$(d). Broad Gaussian beam centered at the edge waveguide is used for array excitation.}
\label{fig:4}
\end{figure}

\section{Bloch oscillation in two-dimensional helical waveguides}

\label{sec:3}

\begin{figure}[htp]
\centering
{\includegraphics[width=1\linewidth]{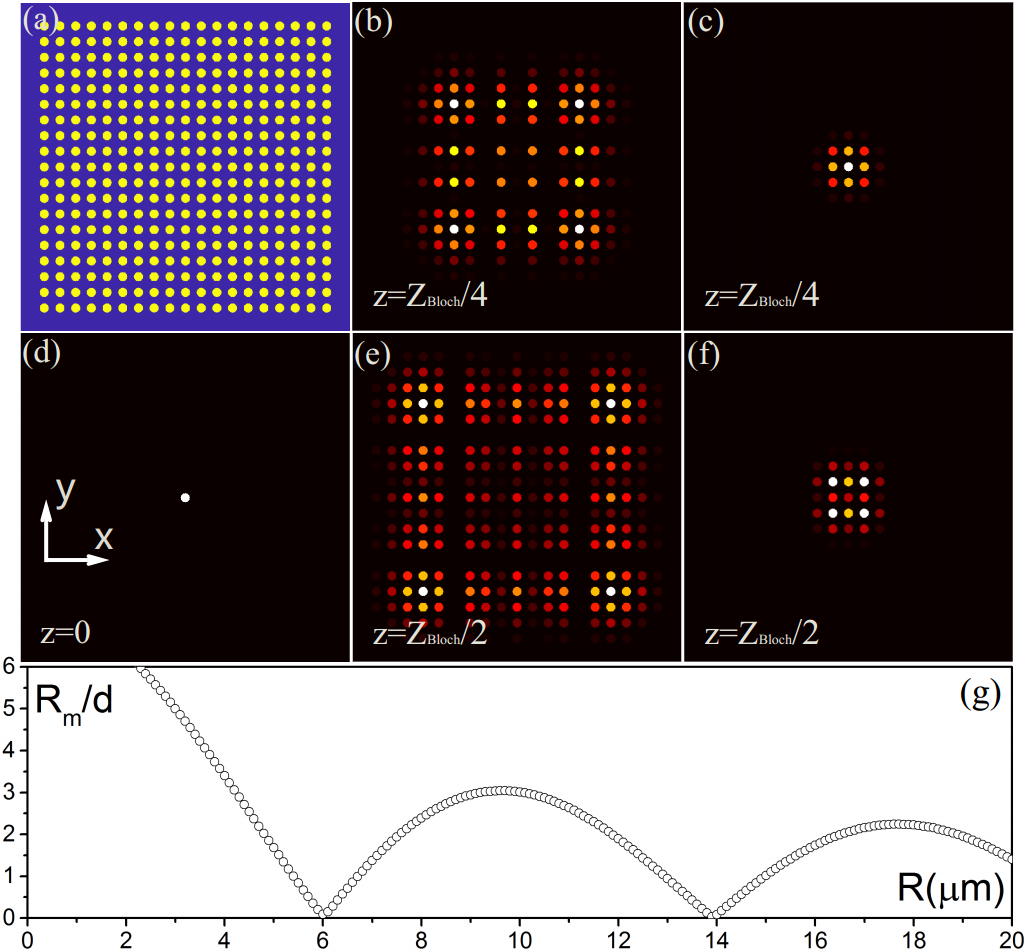}}
\caption{ (a) A sketch of the 2D square waveguide array. Field modulus distributions at $z=Z_\text{Bloch}/4$ (b),(c) and $z=Z_\text{Bloch}/2$ (e),(f) for excitation of only one central waveguide [panel (d)] in the array for helix radius $R=0~\mu \textrm{m}$ (b),(e) and $5~\mu \text{m}$ (c),(f). Maximal scaled radius of the wavepacket $R_\textrm{m}/d$ at $z=Z_\text{Bloch}/2$ as a function of helix radius $R$.}
\label{fig:5}
\end{figure}
The control of BOs dynamics in helical waveguide arrays is readily achievable also in the two-dimensional geometries. Now we consider a square array of helical waveguides [Fig.~5(a)], with a linear refractive index gradient in both $x$ and $y$ directions. The propagation of light in such a structure is described by Eq.~ (1) with the term $\alpha x$ replaced by $\alpha x+\alpha y$. For equal gradients the propagation dynamics in two-dimensional array is qualitatively similar to that in the one-dimensional structure. When only the central waveguide is excited [Fig.~5(d)], the wavepacket first expands in both transverse dimensions and achieves its maximal width at $z=Z_\text{Bloch}/2$ [Figs.~5(b)-5(f)]. The initial distribution is reproduced at $z=Z_\text{Bloch}$. The comparison of patterns for $R=0~\mu \textrm{m}$ [Figs.~5(b),(e)] and $R=5~ \mu \text{m}$ [Figs. 5(c),(f)] cases reveals substantially smaller amplitude of BOs in the array with helical waveguides. To illustrate that BOs strongly depend on helix radius we calculated the dependence of the maximal radius of the wavepacket $R_\textrm{m}/d=[\sum_n\sum_k (n^2+k^2)|\psi_{n,k}|^2)/\sum_n\sum_k|\psi_{n,k}|^2]^{1/2}$ acquired upon propagation on $R$ [Fig. 5(g)]. The arrest of BOs for helix radii corresponding to band collapse is obvious.

\section{Conclusion}

Summarizing, we studied light propagation in the array of helical waveguides with transverse refractive index gradient, both in 1D and 2D geometries. While wavepackets in such modulated systems still experience BOs, their amplitude and direction strongly depend on waveguide rotation radius. Complete arrest and inversion of BOs direction are reported. Our finding suggests potential applications of the helical waveguides in the control of path and direction of light beam propagation. Since both amplitude and direction of Bloch oscillation depend on the helix radius/period or refractive index gradient, the straightforward application is the control of the displacement of the beam in this complex artificial medium. Thus, at quarter of Bloch oscillation cycle for broad beams one can just tune the output displacement by changing refractive index gradient. For critical helix radius/period, the band collapses becoming flat, thus the propagation of any light beam in the structure in this regime is free of diffraction. This implies the application of such arrays for undistorted transmission of various patterns and even images. Among open problems deserving future investigation is the impact of nonlinearity on BOs in helical waveguide arrays.

\section*{Funding Information}

The work of W.Z. and F.Y. is supported by the National Natural Science Foundation of China (61475101;11690033). Y.V.K. acknowledges support from the Severo Ochoa program (SEV-2015-0522) of the Government of Spain, Fundacio Cellex, Generalitat de Catalunya and CERCA.

\end{document}